\documentclass
[aps,prb,twocolumn,floatfix,english,showpacs,10pt]{revtex4-2}%
\usepackage{graphicx}
\usepackage{amsmath}
\usepackage{physics}
\usepackage{amssymb}
\usepackage{colordvi}
\usepackage{verbatim}
\usepackage{xcolor}
\usepackage{mathrsfs}
\usepackage{epsfig}
\usepackage{lipsum}
\usepackage{amsfonts}
\usepackage[unicode=true, breaklinks=false, pdfborder={0 0 1}, backref=false,
colorlinks=true, linkcolor=blue, urlcolor=blue, citecolor=blue]{hyperref}%
\providecommand{\U}[1]{\protect\rule{.1in}{.1in}}
\setcitestyle{numbers,square}
\begin{document}

\title{Chiral locking of magnon flow and electron spin accumulation in their near-field radiative spin transfer}
\author{Xi-Han Zhou}
\affiliation{School of Physics, Huazhong University of Science and Technology, Wuhan 430074, China}
\author{Xiyin Ye}
\affiliation{School of Physics, Huazhong University of Science and Technology, Wuhan 430074, China}
\author{Tao Yu}
\email{taoyuphy@hust.edu.cn}
\affiliation{School of Physics, Huazhong University of Science and Technology, Wuhan 430074, China}

\date{\today}

\begin{abstract}
We report a non-contact mechanism for directional injection of magnons in magnetic films when driven by a spin accumulation $\pmb{\mu}_s$ of electrons of a nearby metallic layer, governed by the long-range dipolar coupling between magnons and electron spins, which spontaneously generates a magnon current ${\bf J}_m$ flowing in the film plane. Crucially, in such near-field radiative spin transfer, the magnon flow ${\bf J}_m$ is always perpendicular to the spin accumulation $\pmb{\mu}_s$, showing a universal chiral locking relation. The spin injection is efficient even when $\pmb{\mu}_s$ is normal to the magnetization, a feature breaking the limitation of the spin transfer by contact exchange interaction.  Our findings reveal the critical role of dipolar chirality in driving the magnon thermal current and paving the way for the functional design of magnonic devices based on near-field radiative spin transfer.
	
\end{abstract}

\maketitle

\section{Introduction}

Near-field radiative heat transfer refers to the transport of thermal energy by electromagnetic radiation between objects separated by less than a thermal wavelength~\cite{heat}, which involves tunneling processes that can lead to significant deviations from Planck's Law for blackbody radiation. The analogous near-field radiative spin transfer is much less exploited since the mutual interaction between electron spins and magnetic materials mediated by electromagnetic radiation is believed to be weaker than the contact exchange interaction~\cite{magnonics1,magnonics3,manchon2019current,chumak2015magnon,demidov2017magnetization,bauer2012spin,chirality}.

Nevertheless, recent experiments show evidence that the long-range dipolar field~\cite{near_field_pumping} may mediate the spin transfer over micron distances between two parallel and electrically insulated ferromagnetic strips deposited on top of a diamagnetic substrate~\cite{Schlitz}.  As a possible pathway, the electric current drives the spin accumulation ${\pmb \mu}_s$ polarized along the magnetization ${\bf M}_{\rm inj}$ in the injector (and ${\bf M}_{\rm det}$ in the detector)  via the anomalous spin Hall effect; the spin accumulation then drives the magnon current in the detector ferromagnet, which is converted to the voltage signal $V_{\rm det}$ via the inverse charge-to-spin current conversion processes. The long-range spin injection is efficient when the magnon current ${\bf J}_m$ flowing in the detector plane is normal to the spin accumulation ${\pmb \mu}_s\parallel {\bf M}_{\rm inj}\parallel {\bf M}_{\rm det}$ that is normal to the injector plane but vanishes when they are parallel, implying a simple relation for voltage $V_{\rm det}\propto ({\bf J}_m\times {\pmb \mu}_s)$.  As an inverse process, Baumgaertl and Grundler report that the dipolar field of magnon current in the yttrium iron garnet (YIG) film switches the \textit{non-contact} nanowire magnetization~\cite{Baumgaertl,Baumgaertl_2}.

\begin{figure}[ht]
	\begin{center}
	{\includegraphics[width=1.0\linewidth]{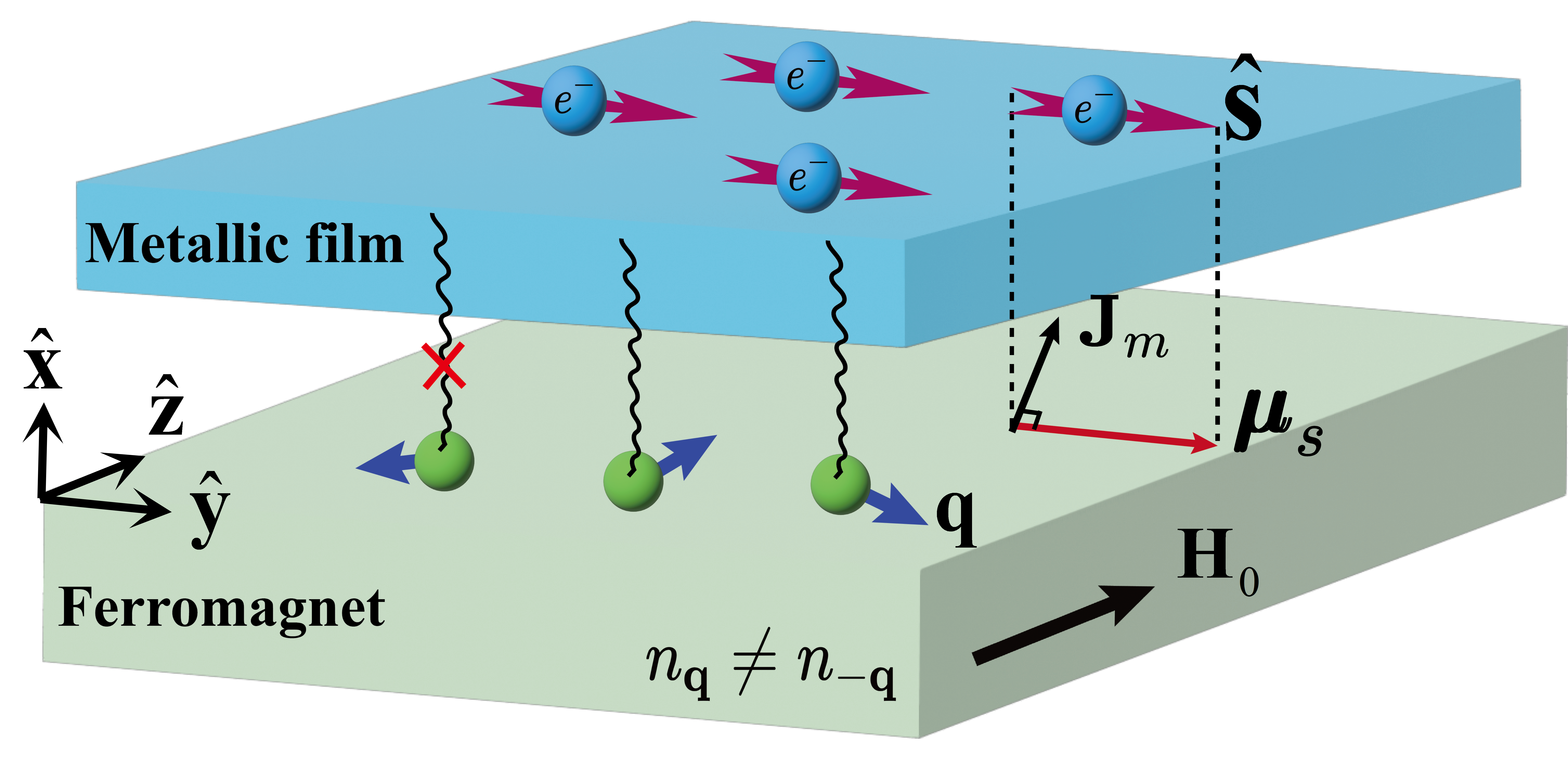}}
\caption{Configuration for the chiral injection of magnons with $n_{\bf q}\ne n_{-{\bf q}}$ that generates the magnon current ${\bf J}_m$ flowing in the ferromagnet plane by nearby spin accumulations ${\pmb \mu}_s$ in metals. The red arrows indicate the spin polarization ${\pmb \mu}_s$ of electrons, while the blue arrows represent the magnon wave vector ${\bf q}$.  An in-plane static magnetic field $H_0\hat{\bf z}$ is applied.}
	\label{model}
	\end{center}
\end{figure}

In this work, we suggest an efficient mechanism for the long-range spin transfer of electron spins to magnon current in a bilayer geometry composed of a heavy metal (or metallic ferromagnet) and a ferromagnet, as illustrated in Fig.~\ref{model}.
We find via the long-range dipolar interaction a spin accumulation $\pmb{\mu}_s$ of electrons in a nearby metallic layer can inject magnons in the magnetic film with asymmetric population $n_{\bf q}\ne n_{-{\bf q}}$ in the Brillouin zone, which spontaneously generates a magnon current ${\bf J}_m$ flowing in the film plane. Crucially, in such a near-field radiative spin transfer, the magnon spin flow ${\bf J}_m$ is perpendicular to the spin accumulation $\pmb{\mu}_s$ of electrons, showing a universal chiral locking relation. Several implications follow: i) When ${\pmb \mu}_s$ is generated by the spin-Hall effect in heavy metals, the generated magnon current ${\bf J}_m$ is \textit{along the applied electric field}, a configuration to date escaping the measurements in the non-local magnon transport~\cite{non_local_transport_1,non_local_transport_2,non_local_transport_3,non_local_transport_4}. ii) The spin injection is still efficient when ${\pmb \mu}_s$ is normal to the magnetization, a feature breaking the limitation of the electric spin injection~\cite{spin_transfer_torque_1,spin_transfer_torque_2,spin_transfer_torque_3} by adjacent exchange interaction.  
iii) When ${\pmb \mu}_s$ is generated by the anomalous spin-Hall effect in a ferromagnetic metal, the chiral locking relation implies the detected voltage $V_{\rm det}\propto ({\bf J}_m\times {\pmb \mu}_s)$ that explains the observation~\cite{Schlitz}. Our findings reveal the critical role of dipolar chirality in driving the magnon thermal current, which can be applied in systems without sufficient exchange transparency for spin transfer.

\section{General formalism}

To be specific, we consider a heterostructure composed of a thin magnetic film of thickness $d$, such as CoFeB, Co, Ni, or YIG, and a thin metallic film of thickness $s$, such as heavy metals that allow a spin accumulation of electrons, as illustrated in Fig.~\ref{model}. The saturation magnetization ${\bf M}_s$ along the $\hat{\bf z}$-direction lies within the magnetic film, so the static interlayer dipolar interaction vanishes. The metal thickness $s$ is smaller than the microwave skin depth $\lambda$, such that the dipolar field ${\bf h}_d$ of magnons could enter inside the metals and affect the dynamics of electrons and vice versa~\cite{Yu_dipolar}. Previous theoretical studies focus on the non-local dipolar interaction between magnons in different magnets~\cite{dipolar_1,dipolar_2} and address the eddy current in metals caused by the dipolar field emitted by the magnon and its backaction~\cite{dipolar_3,dipolar_4}, but the dipolar interaction with the spin degree of freedom of electron has not yet been formulated.
With the spin operators $\hat{\bf s}({\bf r})$ of electrons in the conductor and $\hat{\bf S}({\bf r})$  of magnons in the ferromagnet, the interlayer dipolar coupling reads~\cite{Landau}
\begin{align}
\nonumber
\hat{H}_{\rm int}&=\mu_0\gamma_e\hbar\int_0^s d{\bf r}\hat{\bf s}({\bf r},t)\cdot \hat{{\bf h}}_d({\bf r},t)\\
&=-\frac{\mu_0\gamma_e\gamma\hbar^2}{4\pi}\int_0^s d{\bf r}\hat{{\bf s}}_{\beta}({\bf r})\partial_{\beta}\partial_{\alpha}\int_{-d}^0 d{\bf r}'\frac{\hat{{\bf S}}_{\alpha}({\bf r}')}{|{\bf r}-{\bf r}'|},
\label{dipolar}
\end{align}
where the summation convention is applied over the repeated spatial (or spin) indices $\{\alpha,\beta\}=\{x,y,z\}$.
Here $\mu_0$ is the vacuum permeability, $-\gamma$ is the gyromagnetic ratio
of the magnetic film, and $\gamma_e=|g_e|\mu_B/\hbar$ with the Bohr magneton $\mu_B$ and effective $g$-factor $g_e$ of electrons.

The electron field operator with spin $\xi$ is expanded by the electron wavefunction and annihilation operator $\hat{f}_{{\bf k},\xi}$ as $\hat{\Psi}_{\xi}({\bf r})=({1}/{\sqrt{{\cal A}s}})\sum_{\bf k}e^{i{\bf k  }\cdot{\bf r}}\hat{f}_{{\bf k},\xi}$,
where ${\cal A}$ is  the area of the sample,  ${\pmb \sigma}$ is the Pauli matrices, and ${\pmb \rho}=y\hat{\bf y}+z\hat{\bf z}$, with which the spin operator of electrons 
$\hat{\bf s}({\bf r})=({1}/{2})\hat{\Psi}_{\delta}^{\dagger}({\bf r}){\pmb \sigma}_{\delta\xi}\hat{\Psi}_{\xi}({\bf r})$.
The spin operators in magnetic films are expanded in terms of the magnon operator $\hat{\alpha}_{{\bf q}}$ and magnon ``wavefunction"  $m_{x,y}^{{\bf q}}(x)$  across the film~\cite{Yu1} with the in-plane wave vector ${\bf q}=q_y\hat{\bf y}+q_z\hat{\bf z}$:
\begin{align}
\nonumber
\hat{S}_{x,y}({\bf r})&=\sqrt{2S}\sum_{{\bf q}}\left(m_{x,y}^{{\bf q}}(x)e^{i{\bf q}\cdot{\pmb \rho}}\hat{\alpha}_{{\bf q}}+{m_{x,y}^{{\bf q}*}(x)}e^{-i{\bf q}\cdot{\pmb \rho}}\hat{\alpha}^{\dagger}_{{\bf q}}\right),\\
\hat{S}_z({\bf r})&=-S+(\hat{S}_x^2+\hat{S}_y^2)/(2S),
\label{magnon_operator}
\end{align}
where $S={M_s}/(\gamma \hbar)$.
In the linear-response regime, $\hat{S}_z({\bf r},t)$ does not provide the dipolar electron-magnon interaction. 
Substitution of Eq.~(\ref{magnon_operator}) into (\ref{dipolar}) yields the Hamiltonian of the electron-magnon interaction
\begin{align}
\hat{H}_{\rm int}=\hbar\sum_{{\bf k}{\bf q}}\left({\bf g}({\bf q})\cdot {\pmb \sigma} \right)_{\delta\xi}\hat{f}_{{\bf k}+{\bf q},\delta}^{\dagger}\hat{f}_{{\bf k},\xi}\hat{\alpha}_{\bf q}+{\rm H.c.},
\end{align}
in which the spin-dependent scattering potential
\begin{align}
\nonumber
&\left(\begin{array}{c}
     g_x({\bf q})  \\
     g_y({\bf q})\\
     g_z({\bf q})
\end{array}\right)=\eta F(|{\bf q}|)\left(\begin{array}{c}
    1   \\
    -iq_y/|{\bf q}|\\
    -iq_z/|{\bf q}| 
\end{array}\right)\\
&\times\int_{-d}^0 dx'e^{|{\bf q}|x'}\left(|{\bf q}|m_x^{\bf q}(x')-iq_ym_y^{\bf q}(x') \right),
\label{potentials}
\end{align}
where $\eta\equiv -({\mu_0\gamma_e}/{4})\sqrt{{2M_s}{\gamma\hbar}}$ and the form factor 
\begin{align}
F(|{\bf q}|)=\frac{1}{s}\int_{0}^sdx e^{-|{\bf q}|x}= \frac{1}{|{\bf q}|s}\left(1-e^{-|{\bf q}|s}\right)
\end{align}
takes the average of the dipolar field across the metallic film, which is close to unity and becomes wave-vector independent when $|{\bf q}|s\ll 1$. 
The scattering potential ${\bf g}({\bf q})$ only depends on the wave vector ${\bf q}$ of magnons, analogous to that of the electron-phonon interaction~\cite{mahan}. $g_x({\bf q})$ and $g_y({\bf q})$ induce the spin-flip scattering for electrons, while $g_z({\bf q})$ causes spin-conserved scattering.

For thin magnetic films, the Damon-Eshbach and backward moving spin waves merge into perpendicular standing spin waves (PSSWs)~\cite{PSSW}.
For the dipolar interaction, only the lowest
PSSW contributes. With the frequency 
\begin{align}
\Omega_{\bf q}/\hbar&=\mu_{0}\gamma \bigg{[}(H_0+M_s+\alpha
_{\mathrm{ex}}M_sq^{2})(H_0+\alpha_{\mathrm{ex}}M_sq^{2})\nonumber\\
&+M_s^2(  1+\frac{e^{-|q|d}-1}{|q|d})(\frac{1-e^{-|q|d}}{|q|d})\bigg{]}^{1/2} 
\end{align}
contributed by both dipolar and exchange couplings~\cite{dipolar_exchange}, where $\alpha_{\rm ex}$ is the exchange stiffness, the magnon Hamiltonian $\hat{H}_m=\sum_{\bf q}\Omega_{\bf q} \hat{\alpha}_{\bf q}^\dagger\hat{\alpha}_{\bf q}$. The lowest PSSW  is circularly polarized with $m_y=im_x\rightarrow \sqrt{1/(4{\cal A}d)}$, so the scattering potentials (\ref{potentials}) become   
\begin{align}
\nonumber
g_{x}({\bf q})&\rightarrow -i\sqrt{1/(4{\cal A}d)}\eta F(|{\bf q}|)(1-e^{-|{\bf q}|d})(1+\cos\theta_{\bf q}),\\
g_{y}({\bf q})&\rightarrow -i\cos\theta_{\bf q}g_x({\bf q}),\nonumber\\
g_{z}({\bf q})&\rightarrow -i\sin\theta_{\bf q}g_x({\bf q}).
\label{coupling_constants}
\end{align}
Here, $\cos\theta_{\bf q}=q_y/|{\bf q}|$ and $\sin\theta_{\bf q}=q_z/|{\bf q}|$. 
The scattering potential is enhanced with thin films~\cite{Xiang_yang}.
The coupling strongly depends on the propagation direction of magnons, which is chiral with different strengths when $\theta_{\bf q}=0$ and $\pi$. 

The chirality in the scattering potential ${\bf g}({\bf q})$ originates from the spin-momentum locking of the dipolar field emitted by spin waves propagating normal and parallel to the saturation magnetization ${\bf M}_s\parallel \hat{\bf z}$.

We plot a snapshot of the magnetic moment (black arrows) distribution of spin waves and their dipolar field (red arrows) in Fig.~\ref{spin_wave_field}.  For the Damon-Eshbach configuration ${\bf k}\perp{\bf M}_s$, by adding the stray field of nearby magnetic moments,  only the spin waves propagating to the right ($+\hat{\bf y}$-direction) emits the stray field above the magnetic film to couple the electron spin of the metal, as shown in Fig.~\ref{spin_wave_field}(a). When the spin wave propagates to the left ($-\hat{\bf y}$-direction), the dipolar field of nearby magnetic moment cancels above the film, as shown in Fig.~\ref{spin_wave_field}(b). Moreover, for the spin wave propagating along $+\hat{\bf y}$-direction, the stray field is circularly polarized with circular polarization $\hat{\bf S}_T$ along the $-\hat{\bf z}$-direction. Thereby, only the spin waves propagating along $+\hat{\bf y}$ direction couple with the heavy metal; when the spin accumulation ${\pmb \mu}_s $ is antiparallel to  $\hat{\bf S}_T$, the magnon is excited by spin transfer.

In the bulk-volume configuration ${\bf k}\parallel{\bf M}_s$, for the spin wave propagating along the $\pm \hat{\bf z}$-axes, the dipolar field exists both above and below the magnetic film, as plotted in Fig.~\ref{spin_wave_field}(c) and (d). The circular polarization $\hat{\bf S}_T$ of the stray field orients along the $+\hat{\bf y}$ direction when ${\bf k}\parallel +\hat{\bf z}$ and is inverted when the propagation direction is reversed. When ${\pmb \mu}_s\parallel \hat{\bf y}$, the magnon propagating along $\hat{\bf z}$ holds $-\hat{\bf S}_T\parallel {\pmb \mu}_s$, resulting in a net magnon injection; otherwise, it results in a net absorbtion of magnons propagating along $-\hat{\bf z}$. This leads to the locking of the spin accumulation direction and the magnon flow.

\begin{widetext}
 \begin{center}
\begin{figure}[htp!]
    \centering
    \includegraphics[width=0.78\linewidth]{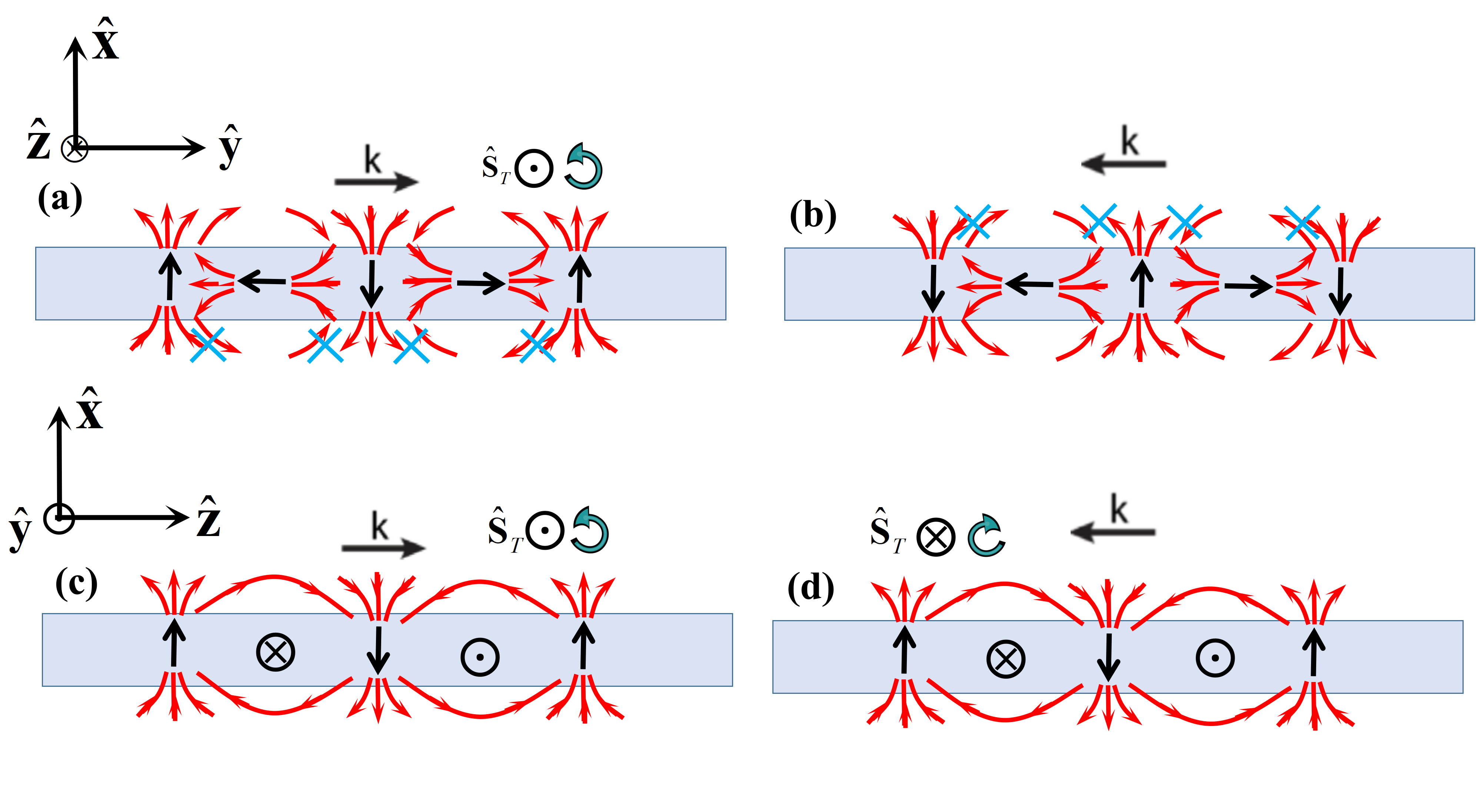}
    \caption{Dipolar-field distribution and its circular polarization $\hat{\bf S}_T$ for the spin waves in the Damon-Eshbach [(a) and (b)] and bulk-volume [(c) and (d)] configurations.  }
    \label{spin_wave_field}
\end{figure}
\end{center}
\end{widetext}

\begin{figure}[ht]
	\begin{center}
       \includegraphics[width=\linewidth]{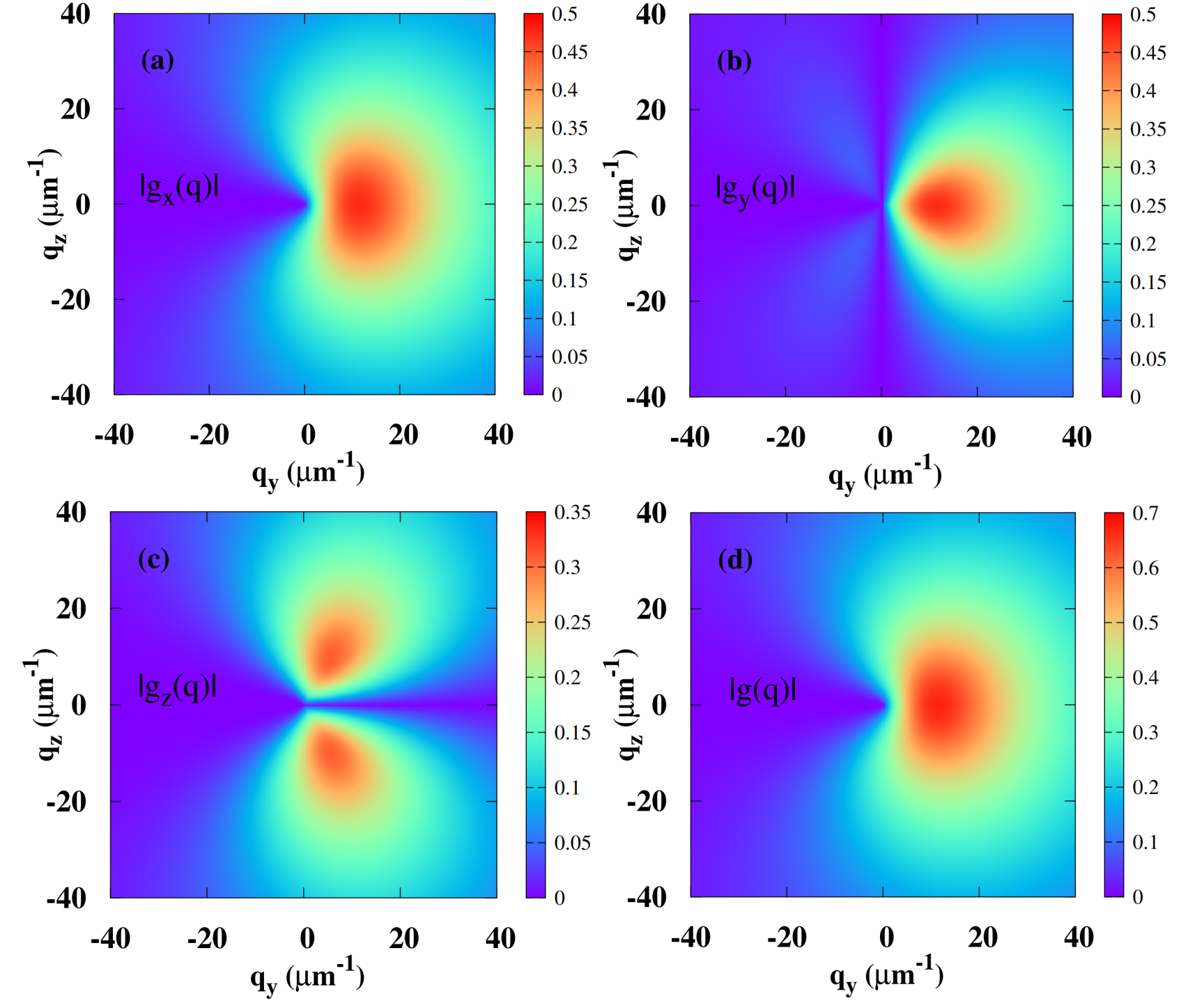}
	\caption{Wave-vector dependence of the interlayer dipolar coupling (unit: $s^{-1}$) between magnons in CoFeB thin film of thickness $d=100$~nm and electron spins in the metallic film of thickness $s=100$~nm. (a)-(d) plots $|g_x({\bf q})|$, $|g_y({\bf q})|$, $|g_z({\bf q})|$, and $|{\bf g}({\bf q})|$, respectively.}
	\label{coupling}
	\end{center}
\end{figure}

We illustrate the wave-vector dependence of the non-local dipolar couplings between the CoFeB thin film of thickness $d=100$~nm with saturation magnetization $\mu_0M_s=1.6$~T and the metallic film of thickness $s=100$~nm in Fig.~\ref{coupling}. 
Only the magnons with positive $q_y$ can efficiently couple to the electron spin. The coupling is not weak for the magnons of wave vectors up to $50$~$\mu$m$^{-1}$ (the wavelength of spin waves is about $100$~nm).

\section{Chiral injection of magnons}

We focus on the injection and transport of incoherent magnons in the magnetic films when biased by the spin accumulation of electrons. The interfacial exchange or Dzyaloshinskii–Moriya interaction is the possible source for the spin-flip scattering that injects magnons by spin accumulation~\cite{Zhang_Zhang,non_local_transport_1}, but it is easily suppressed by inserting a thin insulator, such as Al$_2$O$_3$, while leaving the long-range dipolar interaction less affected~\cite{Bart_thermal,Baumgaertl,Baumgaertl_2,wang2025BiYIG}.

Derived by the Liouville equation, the injection rate of the magnon population  $n_{\bf q}$ reads
\begin{align}
&\frac{\partial n_{\bf q}}{\partial t}\Big|_s={2\pi}{\hbar}\sum_{{\bf k}}\delta(\varepsilon_{{\bf q}+{\bf k}}-\varepsilon_{{\bf k}}-\Omega_{\bf q})\nonumber\\
&\times \Big\{{\rm Tr}\left[\rho_{{\bf q}+{\bf k}}({\bf g}({\bf q})\cdot{\pmb \sigma})(1-\rho_{{\bf k}})({\bf g}({\bf q})\cdot{\pmb \sigma})^{\dagger}\right](1+n_{\bf q})\nonumber\\
&-{\rm Tr}\left[(1-\rho_{{\bf q}+{\bf k}})\left({\bf g}({\bf q})\cdot{\pmb \sigma}\right)\rho_{{\bf k}}\left({\bf g}({\bf q}\right)\cdot{\pmb \sigma})^{\dagger}\right]n_{\bf q}\Big\},
\label{magnon_dy}
\end{align}
where $\varepsilon_{\bf k}$ is the electron dispersion, $\rho_{{\bf k}}$ represents the $2\times 2$
density matrix of electrons of crystal momentum $\hbar{\bf k}$ in the spin space, in
which the diagonal elements $\rho_{s,s}({\bf k},t )$ describe electron
distributions and the off-diagonal elements $\rho_{s,-s}({\bf k},t)$
represent the spin coherence~\cite{Wu_review,Yu_coldatom}.

The driven in-plane electric current in the metallic layer generates the electron spin current injected normal to the film by the spin-Hall effect~\cite{spin_injection_metals,spin_Hall_effect_1,spin_Hall_effect_2,spin_Hall_effect_3,spin_Hall_effect_4}, such that the spin accumulation $\pmb{\mu}_s=\mu_s\hat{\bf n}$ of a decay length $l_{\rm sf}\lesssim 10$~nm polarized in the film plane is strongly localized at two boundaries of thick heavy metals. According to the coupling constant $|{\bf g}({\bf q})|\propto (1/|{\bf q}|)(1-e^{-{|{\bf q}|d}})(1-e^{-|{\bf q}|s})$, the magnons with $|{\bf q}|\sim 1/s\sim 1/d$ is dominantly excited  when $d\sim s$, so their typical decay length $l_m\sim s$ since the stray field decays according to $ e^{-|{\bf q}|x}$. Therefore,  the magnon stray field with decay length $l_m\sim 100$~nm decays about $70\%$ already at the other boundary such that it mainly interacts with the spin accumulation at the interface of heavy metal and ferromagnet. The Oersted field by the charge current in extended heavy-metal layer is homogeneous without trapping magnons~\cite{trapping}. The stray field of magnons causes additional radiative damping $\sim 10^{-5}$ by eddy current in thin heavy metals that is negligible~\cite{dipolar_3,dipolar_4}.

Here, we consider the spin imbalance of electrons with the population of spin ``$\uparrow$'' not equal to that of spin ``$\downarrow$'', governed by different chemical potentials $\mu_\uparrow=\varepsilon_{F}+\mu_s/2$ and $\mu_\downarrow=\varepsilon_{F} -\mu_s/2$ deviating from Fermi energy $\varepsilon_{F}$.  With the polarization direction  $\hat{\bf n}=(n_x,n_y,n_z)$,  the density matrix of electrons
\begin{align}
\rho_{{\bf k}}=({{\cal F}_{\bf k,\uparrow}+{\cal F}_{\bf k,\downarrow}})/{2}{\cal \pmb I}_{2\times 2}+({{\cal F}_{\bf k,\uparrow}-{\cal F}_{\bf k,\downarrow}})/{2}{\pmb \sigma\cdot \hat{\bf n}},
\label{distribution}
\end{align}
where ${\cal{\pmb I}}_{2\times 2}$ is the $2 \times 2$ identity matrix, ${\cal F}_{\bf k,\zeta}=(e^{\beta {\cal E}_{\bf k,\zeta}}+1)^{-1} $ is the Fermi-Dirac distribution with ${\cal E}_{\bf k,\uparrow}=\varepsilon_{\bf k}-\mu_{\uparrow}$ and ${\cal E}_{\bf k,\downarrow}=\varepsilon_{\bf k}-\mu_{\downarrow}$.  With $\mu_s \ll \varepsilon_{F}$, the distribution 
\begin{align}
   &{\cal F}_{\bf k,\uparrow/\downarrow} \approx {\cal F}^{(0)}_{\bf k}\mp({\mu_s}/{2}){\partial {\cal F}_{\bf k}^{(0)}}/{\partial \varepsilon_{\bf k}},\nonumber\\
   &{\cal F}_{{\bf k+q},\uparrow/\downarrow}\approx {\cal F}_{\bf k}^{(0)}-\left(\pm{\mu_s}/{2}-\Omega_{\bf q}\right){\partial {\cal F}_{\bf k}^{(0)}}/{\partial \varepsilon_{\bf k}},
   \label{expand_F}
 \end{align}
where  ${\cal F}_{\bf k}^{(0)}=[\exp(\beta (\varepsilon_{\bf k}-\varepsilon_F))+1]^{-1}$ is the electron equilibrium Fermi-Dirac distribution. 
Substituting Eqs.~\eqref{distribution} and \eqref{expand_F} into \eqref{magnon_dy} yields
\begin{align}
    \frac{\partial n_{\bf q}}{\partial t}\Big|_s=&2\pi \hbar\sum_{\bf k}\delta(\varepsilon_{\bf k+q}-\varepsilon_{\bf k}-\Omega_{\bf q}) \dfrac{\partial {\cal F}_{\bf k}^{(0)}}{\partial \varepsilon_{\bf k}}\left(
     n_{\bf q} {\cal Q}({\bf q})+\tilde{\cal Q}({\bf q})
    \right),
    \nonumber
\end{align}
in which with ${\cal P}_1({\bf q})=g^2_z({\bf q})$, ${\cal P}_2({\bf q})=g^2_y({\bf q})$, and ${\cal P}_3({\bf q})=-g^2_x({\bf q})$, ${\cal Q}({\bf q})=2\Omega_{\bf q}({\cal P}_1({\bf q})+{\cal P}_2({\bf q})+{\cal P}_3({\bf q})) - 4 i \mu_s g_x({\bf q}) (g_y({\bf q})n_z-g_z({\bf q})n_y)$ and  
 \begin{align}
     &\tilde{\cal Q}({\bf q})= -2 g_y({\bf q})g_z({\bf q}) n_yn_z \nonumber\\
     &\times\left(2  \Omega_{\bf q} N(\Omega_{\bf q})-\Omega_{\bf q}^+N(\Omega_{\bf q}^+)-\Omega_{\bf q}^-N(\Omega_{\bf q}^-) \right)\nonumber\\
     &-2ig_x({\bf q})\left(g_z({\bf q})n_y-g_y({\bf q})n_z\right)\left(\Omega_{\bf q}^+ N(\Omega_{\bf q}^+)-\Omega_{\bf q}^-N(\Omega_{\bf q}^-)\right)\nonumber\\
     &-\frac{1}{2}\sum_{j=1,2,3}{\cal P}_j({\bf q}) [
     2\Omega_{\bf q} N(\Omega_{\bf q})(1 + {\cal A}_j\cosh(\beta\mu_s/2))\nonumber\\
     &+ \Omega_{\bf q}^{+} N(\Omega_{\bf q}^{+}) (1  - {\cal A}_{j}e^{\beta\mu_s/2})+ \Omega_{\bf q}^- N(\Omega_{\bf q}^-) (1 - {\cal A}_{j} e^{-\beta\mu_s/2})],
 \nonumber
 \end{align}
where  $\Omega_{\bf q}^\pm=\Omega_{\bf q}\pm \mu_s $,  $N(\Omega_{\bf k})=(\exp(\beta \Omega_{\bf k})-1)^{-1}$ is the equilibrium Planck distribution of magnons, ${\cal A}_1=n_x^2+n_y^2-n_z^2$, 
${\cal A}_2=n_x^2-n_y^2+n_z^2$, and ${\cal A}_3=-n_x^2 + n_y^2 + n_z^2$.

For metals, ${\partial {\cal F}_{\bf k}^{(0)}}/{\partial \varepsilon_{\bf k}}\approx-\delta(\varepsilon_{\bf k}-\varepsilon_{F})$, 
such that 
\[
{\partial n_{\bf q}}/{\partial t}|_s=2\pi \hbar I_{\bf q}\left(n_{\bf q} {\cal  Q}({\bf q})+\tilde{\cal Q}({\bf q})\right),
\]
where
\begin{align}
    I_{\bf q}=&-\sum_{\bf k}\delta(\varepsilon_{\bf k+q}-\varepsilon_{\bf k}-\Omega_{\bf q})\delta(\varepsilon_{\bf k}-\varepsilon_{F})\nonumber\\
    =&-\dfrac{s{\cal A}}{(2 \pi)^2}\dfrac{m^2}{\hbar^4 q}\Theta\left(1-\left(\dfrac{m\Omega_{\bf q}}{\hbar^2 k_F q}-\dfrac{q}{2 k_F}\right)^2\right).
    \label{delta_function}
\end{align}
The injected magnon relaxes to its equilibrium distribution $N_{\bf q}$ within the relaxation time $\tau_{\bf q}$. The injection and relaxation lead to the quantum kinetic equation of magnons 
${\partial n_{\bf q}}/{\partial t}=-({n_{\bf q}-N_{\bf q}})/{\tau_{\bf q}}+2\pi \hbar I_{\bf q}(
     n_{\bf q} {\cal Q}({\bf q})+\tilde{\cal Q}({\bf q})
    )$.
The steady-state solution  
\begin{align}
    n_{\bf q}=\dfrac{N_{\bf q}+2 \pi\hbar I_{\bf q}\tau_{\bf q}\tilde{\cal Q}({\bf q})}{1-2\pi\hbar I_{\bf q}\tau_{\bf q} {\cal Q}({\bf q})}. 
    \label{excited_magnon}
\end{align}
When the spin accumulation $\mu_s$ is sufficiently large, an instability is implied by $2\pi\hbar I_{\bf q}\tau_{\bf q} {\cal Q}({\bf q})\rightarrow 1$. 
We focus on the smaller spin accumulation without inducing the change of magnetic configuration. 
In this case,  $2\pi\hbar I_{\bf q}\tau_{\bf q} {\cal Q}({\bf q})\ll1$ and $2\pi\hbar I_{\bf q}\tau_{\bf q} \tilde{\cal Q}({\bf q})\ll N_{\bf q}$, such that the injected magnons into the magnetic film  
\begin{align}
    \delta n_{\bf q}&\equiv n_{\bf q}-N_{\bf q}\approx 2 \pi\hbar N_{\bf q}I_{\bf q}\tau_{\bf q}{\cal Q}({\bf q})\nonumber\\
    &=2 \pi\hbar N_{\bf q}I_{\bf q}\tau_{\bf q}\left(2 \Omega_{\bf q} |{\bf g}({\bf q})|^2 -4\mu_s g_x^2({\bf q}) \cos{(\theta_{\bf q}+\phi)}\right)\nonumber\\
    &\overset{\mu_s\gg \Omega_{\bf q}}{\approx} -8 \pi\hbar N_{\bf q}I_{\bf q}\tau_{\bf q}\mu_s g_x^2({\bf q}) \cos{(\theta_{\bf q}+\phi)},
    \label{excited_magnon2}
\end{align}
where $\phi$ is the angle of the spin-polarization direction $\hat{\bf n}$ with respect to the saturation magnetization \(\hat{\bf z}\)-direction.
The distribution of the injected magnon $\delta n_{\bf q}$ in the Brillouin zone mainly depends on $g_x^2({\bf q})<0$ [Eq.~\eqref{coupling_constants}], which is asymmetric as indicated by Fig.~\ref{coupling}(a). On the other hand, the direction of the spin polarization with respect to the saturation magnetization direction modulates the spin injection via $\cos{(\theta_{\bf q}+\phi)}$. Since $I_{\bf q}<0$ in Eq.~\eqref{delta_function}, the magnons are excited when  $\cos{(\theta_{\bf q}+\phi)}<0$ with $\pi/2<(\theta_{\bf q}+\phi)<3\pi/2$, while they are absorbed by metals when $\cos{(\theta_{\bf q}+\phi)}>0$ with $-\pi/2<(\theta_{\bf q}+\phi)<\pi/2$.

\begin{figure}[htp!]
    \centering
    \includegraphics[width=1.01\linewidth]{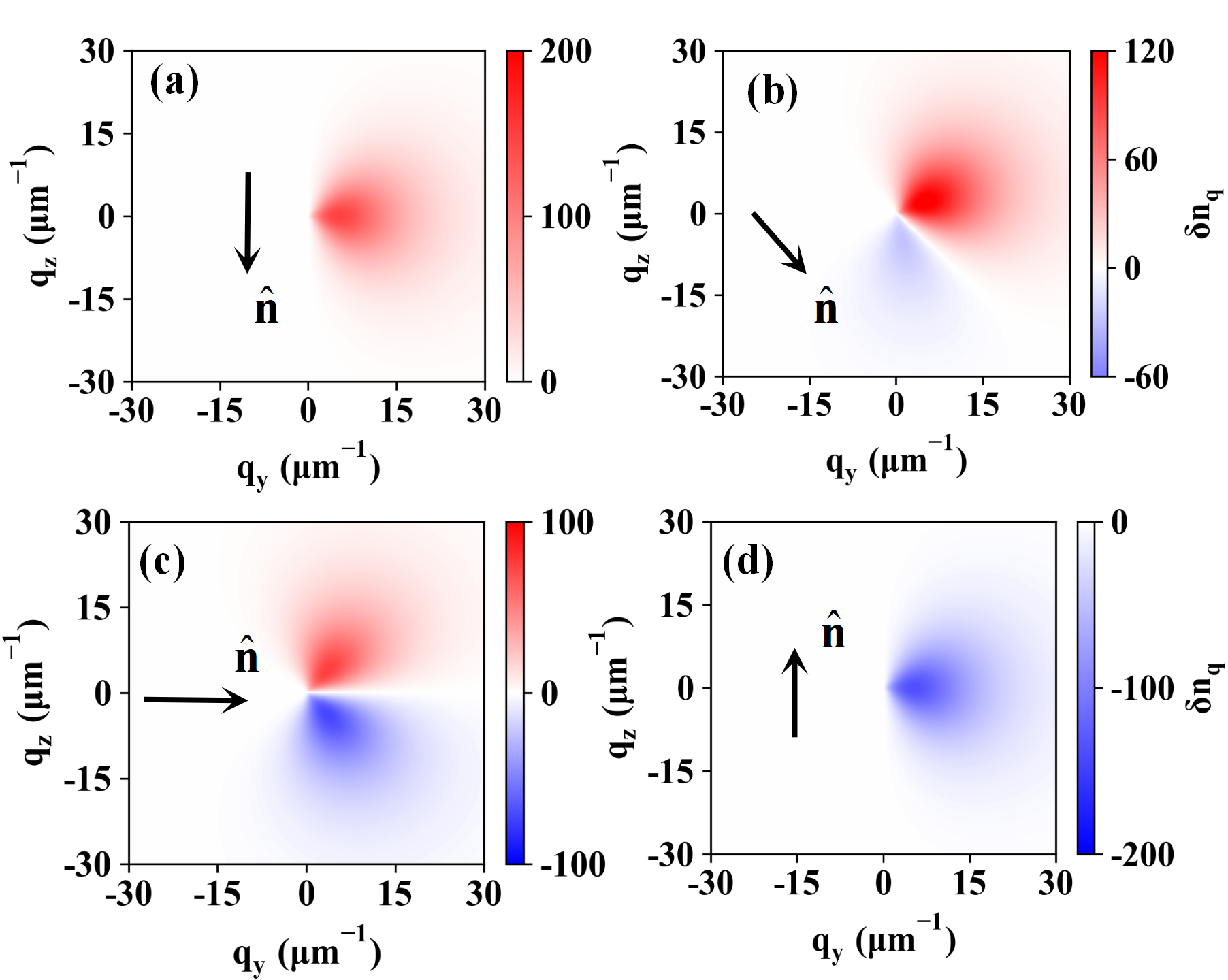}
    \caption{Distribution of injected magnons $\delta n_{\bf q}$ in the Brillouin zone when biased by different spin-accumulation directions $\phi=\pi$ [(a)], $3\pi/4$ [(b)], $\pi/2$ [(c)], and $0$ [(d)].}
    \label{magnon_number}
\end{figure}

To illustrate the physical principle, we consider the typical metals of thickness $s=100$~nm with $\varepsilon_F\approx 3.8$~eV and  $\mu_s=2$~meV~\cite{spin_accumulation_1,spin_accumulation_2} as the spin battery to the CoFeB thin film of thickness $d=100$~nm, saturation magnetization $\mu_0M_s=1.6$~T, exchange stiffness $\alpha_{\rm ex}=1.3\times10^{-17}~{\rm m}^{2}$, and Gilbert damping $\alpha_G=10^{-3}$~\cite{cofeb1,cofeb2,cofeb3}, biased by the static magnetic field $\mu_0 H_0=60$~mT along the $\hat{\bf z}$-direction.  
It leads to the relaxation time $\tau_{\bf q}=(\alpha_{G}\Omega_{\bf q}/\hbar)^{-1}$ and  $\Omega_{\bf q}\sim 0.01$~meV $\ll \mu_s$.  
Figure~\ref{magnon_number} shows the numerical results for the injected magnons $\delta n_{\bf q}$ with different spin-polarization directions $\hat{\bf n}$. Only magnons in the half Brillouin zone with $ q_y>0$ are modulated because of the chirality in the dipolar coupling [Fig.~\ref{coupling}(a)]. 
When the spin polarization is opposite to the saturation magnetization ($\phi=\pi$) as in Fig.~\ref{magnon_number}(a), the coupling leads to the magnon excitation with wave vectors $ q_y>0$ since $\pi/2<(\theta_{\bf q}+\phi)<3\pi/2$, implying a magnon spin flux ${\bf J}_{m}$ along the positive $\hat{\bf y}$-direction, which is perpendicular to $\hat{\bf n}$. 
When the spin-accumulation direction is away from $\phi=\pi$, the magnon absorption occurs, as shown in Fig.~\ref{magnon_number}(b). In particular, when $\hat{\bf n}$ is normal to the saturation magnetization ($\phi=\pi/2$), almost equal excitation and absorption of magnons occur for $0<\theta_{\bf q}<\pi/2$  and $-\pi/2<\theta_{\bf q}<0$, as shown in Fig.~\ref{magnon_number}(c), implying a magnon spin flux ${\bf J}_{m}\perp \hat{\bf n}$ along the saturation magnetization. 
When the spin accumulation in Fig.~\ref{magnon_number}(a) is reversed to along the saturation magnetization ($\phi=0$), as in Fig.~\ref{magnon_number}(d), the dipolar coupling results in the magnon absorption with wave vectors $ q_y>0$, indicating a reversed magnon spin flux ${\bf J}_m$ compared with that in Fig.~\ref{magnon_number}(a).

\begin{figure}
    \centering
    \includegraphics[width=1.01\linewidth]{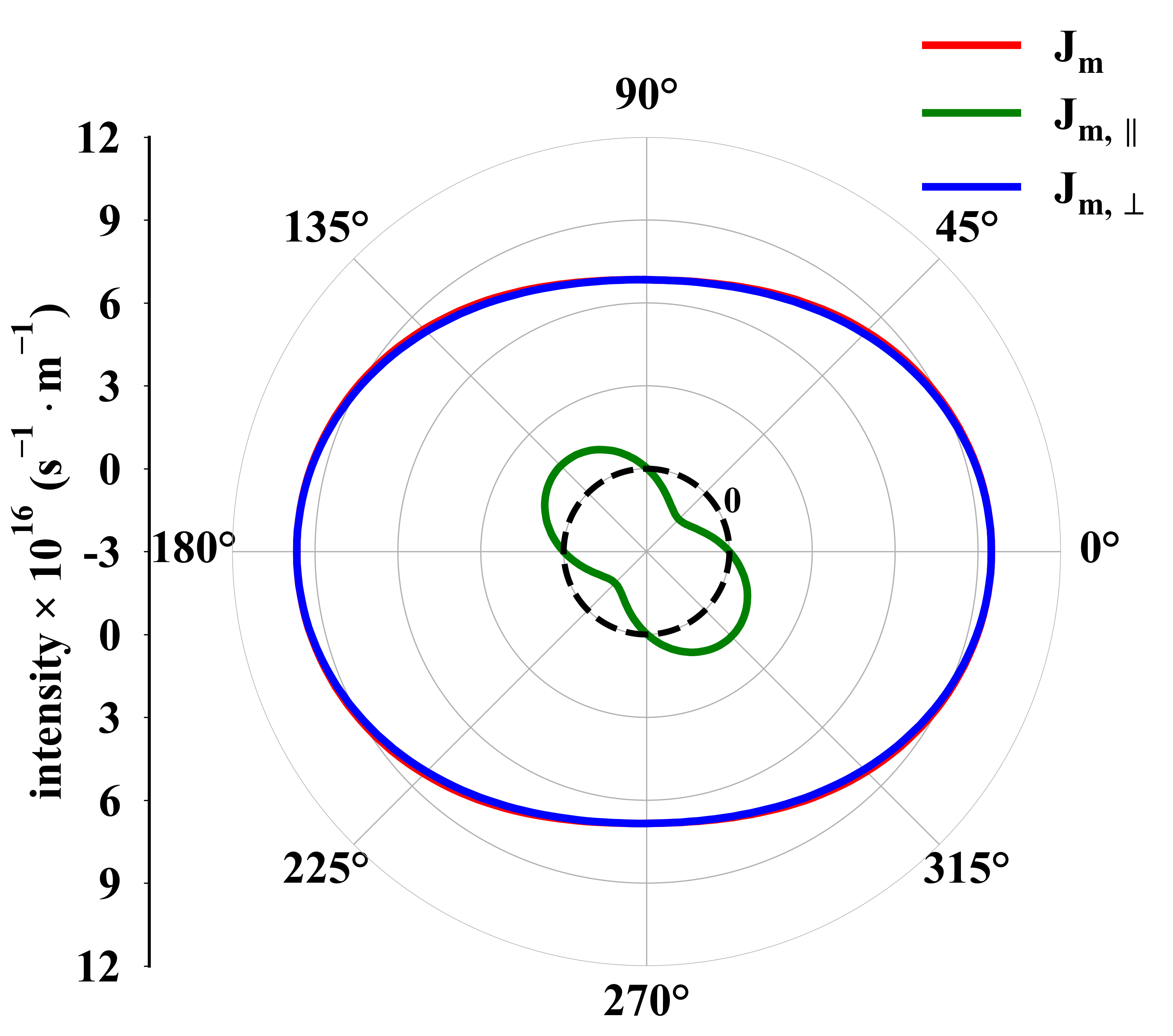}
    \caption{Driven magnon flow ${\bf J}_m$ as a function of the in-plane spin-accumulation directions $\phi$ with respect to the saturation magnetization direction. ${\bf J}_{m,\parallel}$ and ${\bf J}_{m,\perp}$ are the parallel and perpendicular components of ${\bf J}_m$ with respect to the spin-accumulation direction $\hat{\bf n}$.}
    \label{magnon_flux}
\end{figure}

${\bf J}_m=(1/{\cal A}) \sum_{\bf q} {\bf v}_{\bf q} n_{\bf q} \perp \hat{\bf n}$ is a universal feature in the near-field spin transfer, where ${\bf v}_{\bf q}= \partial \Omega_{\bf q}/(\hbar\partial{\bf q})$ is the group velocity of magnons. We find that generally for the parallel and normal components 
\begin{align}
    {\bf J}_{m,\parallel}&\equiv \hat{\bf n}\cdot {\bf J}_{m}\propto\dfrac{1}{2}\int_0^{2\pi} d\theta_{\bf q} \sin{\left(2(\theta_{\bf q}+\phi)\right)}(1+\cos{\theta_{\bf q}})^2\nonumber\\
    &\propto\sin(2\phi),\nonumber\\
    {\bf J}_{m,\perp}&\equiv \hat{\bf n}\times {\bf J}_{m}\propto\int_0^{2\pi} d\theta_{\bf q} (-1)\cos^2{(\theta_{\bf q}+\phi)}(1+\cos{\theta_{\bf q}})^2\nonumber\\
    &\propto-(6+\cos(2\phi)),
\end{align}
\({\bf J}_{m,\parallel}/{\bf J}_{m,\perp}=-\sin{(2\phi)}/(6+\cos(2\phi))\ll 1\). The chiral locking is exact when $\phi=\{0,\pi/2,\pi,3\pi/2\}$, i.e., \({\bf J}_{m,\parallel}/{\bf J}_{m,\perp}=0\). Figure~\ref{magnon_flux} demonstrates above feature by comparing ${\bf J}_{m,\parallel}$ and ${\bf J}_{m,\perp}$ as a function of spin-accumulation directions $\phi$.

\section{Discussion and conclusion}

The dipolar coupling of magnons to electron spin is enhanced by the saturation magnetization, which with large saturation magnetizations (\textit{e.g.}, CoFeB) is not significantly weaker than the realistic exchange interaction~\cite{exchange_splitting_1,exchange_splitting_2}. Soft magnets, such as YIG, have a small ratio of dipolar to exchange coupling. As detailed in Appendix~\ref{appendix_b}, we compare the dipolar electric injection of magnons to CoFeB with that by realistic interfacial exchange interaction~\cite{exchange_splitting_1,exchange_splitting_2} and find that its efficiency achieves tens of percent of that by the exchange interaction, which should be within the experimental resolution. Its chirality brings new functionalities in magnonic devices: i) The dipolar interaction causes the non-reciprocity in the non-local transport of magnons. ii) It induces chiral thermal magnon current in the extended bilayer configuration, absent by the exchange interaction since its injected magnons are symmetric in the wave-vector space~\cite{exchange_injection,Zhang_Zhang}.

In conclusion, we have formulated the near-field radiative spin transfer between magnons and electrons mediated by the long-range dipolar coupling and found a universal chiral locking relation between the injected magnon current and its biased spin accumulation of electrons. Although the non-local magnon transport~\cite{non_local_transport_1,non_local_transport_2,non_local_transport_3,non_local_transport_4} has been performed for one decade, this effect may escape its measurement since the chiral locking relation implies the pumped magnon current is along the electric field or the Pt strip. The inverse process implies that a magnon current can convert to the spin accumulation in the nearby metallic ferromagnet that contributes to magnetization switching~\cite{Baumgaertl}. Therefore, the radiative spin transfer is a new effect that may be exploited widely in spin injection/absorption, spin magnetoresistance, and magnetization switching in future experiments.

\begin{acknowledgments}
This work is financially supported by the National Natural Science Foundation of China under Grant No.~12374109 and the National Key Research and Development Program of China under Grant No.~2023YFA1406600. 
\end{acknowledgments}

\begin{appendix}

\section{Electric injection of magnons by exchange interaction}

    \label{appendix_b}

In this appendix, we compare the efficiency of the dipolar electric injection of magnons with that of the interfacial exchange interaction.

For the interface of area ${\cal A}$  between the heavy metal of thickness $s$ and ferromagnetic film of thickness $d$ with the normal along the $\hat{\bf x}$-direction, the interfacial exchange Hamiltonian between the electron spin $\hat{\bf s}({\bf r})$ and magnetization $\hat{\bf M}({\bf r})$ may be written as 
\begin{align}
    \hat{H}_{\rm ex}&=J \hbar\int d{\bf r}\delta(x)\hat{\bf s}({\bf r})\cdot \hat{\bf M}({\bf r})\nonumber\\
    &=J\hbar \int d{\pmb \rho}\hat{\bf s}({\pmb \rho},x=0)\cdot \hat{\bf M}({\pmb \rho},x=0),
    \label{interfacial_exchange}
\end{align}
due to which the exchange coupling constant $J$ induces the exchange splitting $\Delta=J\hbar M_s/s$. $\Delta$ is measurable in the experiment~\cite{exchange_splitting_1,exchange_splitting_2}, from which one can extract $J$. 
The electrons $\hat{f}_{{\bf k},\xi}$ of wave vector ${\bf k}$ and spin $\xi$ interact with magnons $\hat{\alpha}_{\bf q}$ of wave vector ${\bf q}$ via the interfacial exchange Hamiltonian 
\begin{align}
    \hat{H}_{\rm ex}=\hbar\sum_{{\bf k}{\bf q}}\left({\bf p}({\bf q})\cdot {\pmb \sigma}\right)_{\delta\xi}\hat{f}^{\dagger}_{{\bf k}+{\bf q},\delta}\hat{f}_{{\bf k},\xi}\hat{\alpha}_{\bf q}+{\rm H.c.}, 
    \label{exchange_hamiltonian}
\end{align}
with the coupling constant 
\begin{align}
    {\bf p}({\bf q})=-J\gamma \hbar\frac{1}{4s}\sqrt{\frac{2M_s}{\gamma\hbar}}\sqrt{\frac{1}{{\cal A}d}}\left(1,i,0\right)^T,
\end{align}
in which $M_s$ is the saturation magnetization of the magnetic film along the $\hat{\bf z}$-direction. 
The ratio of dipolar to exchange interaction
\begin{align}
    {|{\bf g}({\bf q})|}/{|{\bf p}({\bf q})|}={\mu_0\gamma s}/{J}={\mu_0\gamma \hbar M_s}/{\Delta}.
    \label{ratio}
\end{align}
According to Refs.~\cite{exchange_splitting_1,exchange_splitting_2}, $\Delta\sim 1$~meV for EuS ($\mu_0M_s\sim 0.6$~T) interfacing with Cu (or Al) with thickness $s\sim 5$~nm. So $\mu_0\gamma \hbar M_s=0.2$~meV with CoFeB ($\mu_0M_s=1.6$~T) indicates that the dipolar coupling constant is not so smaller than the exchange one. On the other hand, the exchange coupling is suppressed with the heavy-metal thickness in that $\Delta\propto 1/s$ (\textit{e.g.}, $\Delta\sim 0.05$~meV when $s=100$~nm), while the dipolar interaction is not so sensitive to the heavy metal thickness.

The exchange interaction \eqref{exchange_hamiltonian} injects magnon population $n_{\bf q}$ governed by the injection rate 
\begin{align}
\frac{\partial n_{\bf q}}{\partial t}\Big|_s&\approx {2\pi}{\hbar} n_{\bf q} \sum_{{\bf k}}\delta(\varepsilon_{{\bf q}+{\bf k}}-\varepsilon_{{\bf k}}-\Omega_{\bf q}) \nonumber\\
&\times\Big\{{\rm Tr}\left[\rho_{{\bf q}+{\bf k}}({\bf p}({\bf q})\cdot{\pmb \sigma})(1-\rho_{{\bf k}})({\bf p}({\bf q})\cdot{\pmb \sigma})^{\dagger}\right]\nonumber\\
&-{\rm Tr}\left[(1-\rho_{{\bf q}+{\bf k}})\left({\bf p}({\bf q})\cdot{\pmb \sigma}\right)\rho_{{\bf k}}\left({\bf p}({\bf q}\right)\cdot{\pmb \sigma})^{\dagger}\right]\Big\}\nonumber\\
&\approx -   {8 i\pi \hbar I_{\bf q}\tau_{\bf q} p_x({\bf q}) p_y({\bf q}) \mu_s n_z}n_{\bf q},
\label{magnon_dy}
\end{align}
in which $\varepsilon_{\bf k}$ and $\Omega_{\bf q}$ are the dispersion of electrons in metals and magnons, $\rho_{\bf k}$ is the $2\times 2$ density matrix of electrons, and $n_z$ is the projection $z$-component of spin accumulation ${\pmb \mu}_s$ direction to the saturation magnetization $M_s\hat{\bf z}$. 
On the other hand, the injected magnons relax to equilibrium population $N_{\bf q}$ in the characteristic time $\tau_{\bf q}$. Combining with \eqref{magnon_dy} the magnon dynamics is governed by  
\begin{align}
    \dfrac{\partial n_{\bf q}}{\partial t}=&-\dfrac{n_{\bf q}-N_{\bf q}}{\tau_{\bf q}}- {8 i\pi \hbar I_{\bf q}\tau_{\bf q} p_x({\bf q}) p_y({\bf q}) \mu_s n_z}n_{\bf q}.
\end{align} 
In the steady state, the excited or absorbed magnons
\begin{align}
    \delta n_{\bf q}\approx -   {8 i\pi \hbar I_{\bf q}\tau_{\bf q} p_x({\bf q}) p_y({\bf q}) \mu_s n_z}N_{\bf q},
    \label{delta_nq}
\end{align}
in which
\begin{align}
    I_{\bf q}&=-\sum_{\bf k}\delta(\varepsilon_{\bf k+q}-\varepsilon_{\bf k}-\Omega_{\bf q})\delta(\varepsilon_{\bf k}-\varepsilon_{F})\nonumber\\
    &=\dfrac{-s{\cal A}}{(2 \pi)^2}\dfrac{m^2}{\hbar^4 q}\Theta\left(1-\left(\dfrac{m\Omega_{\bf q}}{\hbar^2 k_F q}-\dfrac{q}{2 k_F}\right)^2\right)
    \label{delta_function}
\end{align}
restricts that only magnons with wave vector
$|\frac{m\Omega_{\bf q}}{\hbar^2 k_F q}-\frac{q}{2 k_F}|<1$
are excited, where $m$ is the mass of electron, and $\varepsilon_F$ and $k_F$ are the Fermi energy and Fermi wave vector. In particular, when the spin accumulation direction is perpendicular to the magnetization direction, i.e., $n_z=0$, the magnon injection vanishes, different from the dipolar thermal pumping.

With $\Delta\sim 0.1$~meV when $s\sim 40$~nm and the same other parameters in the main text, we plot the excited magnon $\delta n_{\bf q}$ by the exchange interaction for different spin-accumulation directions $\hat{\bf n}$ in Fig.~\ref{magnon_exchange}. 
According to Eqs.~\eqref{delta_nq} and \eqref{delta_function}, since ${\bf p}({\bf q})$ is contant, the distribution of $\delta n_{\bf q}$ in the wave-vector space is mainly governed by the term $I_{\bf q}\propto 1/q$, which is large when $q$ is small.
Therefore, we find the exchange interaction mainly pumps the magnons with small wave vectors, as shown in Fig.~\ref{magnon_exchange}. 
This is in contrast to the pumped magnons \begin{align}
    \delta n_{\bf q}^{\rm dipolar}\approx -8 \pi\hbar N_{\bf q}I_{\bf q}\tau_{\bf q}\mu_s g_x^2({\bf q}) \cos{(\theta_{\bf q}+\phi)}
 \end{align}
 by the dipolar interaction, where the coupling due to the dipolar interaction $g_{x}({\bf q})\rightarrow -i\sqrt{1/(4{\cal A}d)}\eta F(|{\bf q}|)(1-e^{-|{\bf q}|d})(1+\cos\theta_{\bf q})$.
 The term $(1-e^{-|{\bf q}|d})\sim qd $ and $g^2_{x}({\bf q})\sim q^2 $, such that $I_{\bf q} g^2_x({\bf q}) \sim q$ when $q$ is small, so no pumped magnons when $q\rightarrow 0$ by the dipolar interaction. 

 \begin{widetext}
 \begin{center}
 \begin{figure}[htp!]
    \centering
    \includegraphics[width=0.95\linewidth]{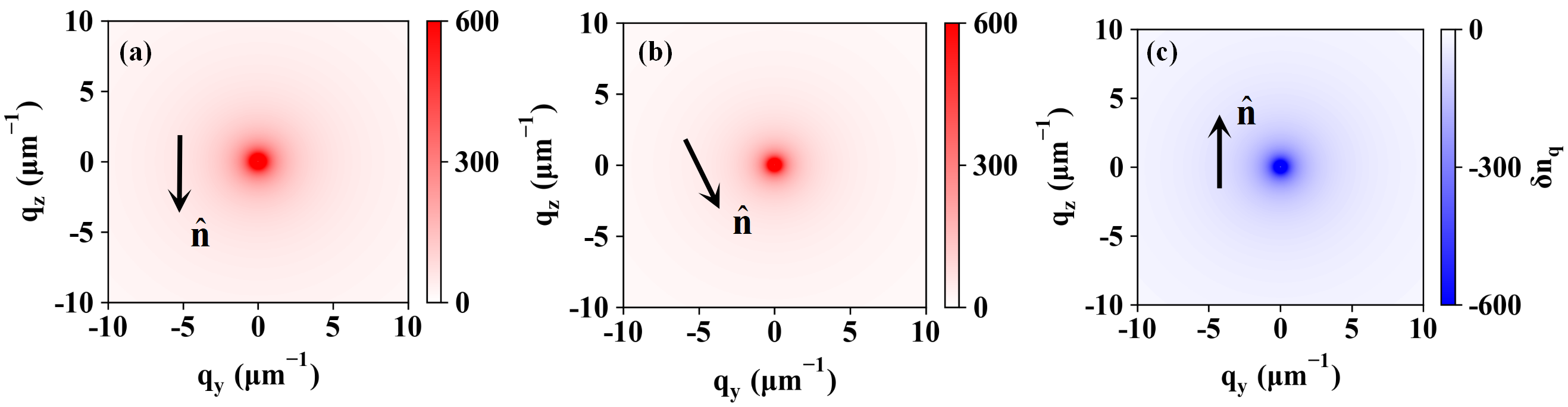}
    \caption{The distribution of injected magnons $\delta n_{\bf q}$ caused by the exchange interaction for different spin-accumulation directions $\phi=\pi$ [(a)], $3\pi/4$ [(b)], and $0$ [(c)] with respect to the saturation magnetization $\hat{\bf z}$-direction. }
    \label{magnon_exchange}
\end{figure}
\end{center}
\end{widetext}

Since no magnon spin current is generated in the bilayer configuration via the exchange interaction, we compare the efficiency of the dipolar and exchange interactions by comparing the total pumped magnon number $\tilde{N}=\sum_{\bf q} \delta n_{\bf q}$. When $n_z=-1$, $\tilde{N}_{\rm dipolar}\sim 3.4\times 10^6$ by the dipolar interaction and $\tilde{N}_{\rm exchange}\sim 16\times 10^6$ by the exchange interaction, so the efficiency of the exchange interaction is only several times in magnitude larger than the dipolar interaction. The exchange interaction becomes less efficient with thicker heavy metals.

\end{appendix}

\end{document}